\documentstyle[12pt,psfig]{article}
\begin{document}
\begin{titlepage}
\begin{center}

{\Large\bf{Seeking the Shadowing in $eA$ Processes}}
\\[5.0ex]
{\Large\it{ M. B. Gay  Ducati $^{*}$\footnotetext{$^{*}$E-mail:gay@if.ufrgs.br}}}\\
 {\it and}\\
{ \Large \it{ V. P.  Gon\c{c}alves $^{**}$\footnotetext{$^{**}$E-mail:barros@if.ufrgs.br} 
}} \\[1.5ex]
{\it Instituto de F\'{\i}sica, Univ. Federal do Rio Grande do Sul}\\
{\it Caixa Postal 15051, 91501-970 Porto Alegre, RS, BRAZIL}\\[5.0ex]
\end{center}

{\large \bf Abstract:}
We consider the behavior of the slope of the nuclear structure function $F_2^A$ in the kinematic region which will be explored in the $eA$ HERA collider. We demonstrate that, similarly to the nucleon case, a 'turn over' is predicted in this observable. Moreover, we predict that the presence of  the shadowing corrections  implies that the maximum value of the slope is dependent of the number of nucleons $A$, differently from the DGLAP predictions. Our conclusion is that the measurement of this observable will allow to evidentiate the presence of the shadowing corrections.

\vspace{1.5cm}



\end{titlepage}

\section{Introduction}
\label{int}

In recent years several experiments have been dedicated to high precision measurements of deep inelastic lepton scattering (DIS) off nuclei. Experiments at CERN and Fermilab focus especially on the region of small values of the Bjorken variable $x = Q^2/2M\nu$, where $Q^2=-q^2$ is the squared four-momentum transfer, $\nu$ the energy transfer and $M$ the nucleon mass.
The data \cite{arneodo}, taken over a wide kinematic range $10^{-5}\,\le\,x\le\,0.1$ and  $0.05\,GeV^2\,\le\,Q^2\le\,100\,GeV^2$, show a systematic reduction of the nuclear structure function $F_2^A(x,Q^2)/A$ with respect to  the free nucleon structure  function $F_2^N(x,Q^2)$. This phenomena is known as {\it the shadowing effect}. The analysis of the shadowing corrections for the nuclear case in deep inelastic scattering (DIS) has been extensively discussed \cite{hera96}. It is motivated by the perspective that in a near future an experimental investigation of the nuclear shadowing at small $x$ and $Q^2 >> 1 \, GeV^2$ using $eA$ scattering could occur at DESY Hadron Electron Ring Accelerator (HERA). Measurements over the extended $x$ and $Q^2$ ranges, which would become 
possible at HERA, will give more information in order to discriminate between the distinct models of shadowing  
and the understanding of the phenomenon which limits the rise of the proton
structure function $F_2$ at small $x$.

The deep inelastic scattering off a nucleus is usually interpreted in a frame where the nucleus is going very fast. In this case the nuclear shadowing is a result of an overlap in the longitudinal direction of the parton clouds originated from different bound nucleons \cite{qiu}. It corresponds to the 
fact that small $x$ partons cannot be localized longitudinally to better
than the size  of the nucleus. Thus low $x$ partons from different
nucleons overlap spatially creating much larger parton densities than in 
the free nucleon case. This leads to a large amplification of the nonlinear
effects expected in QCD at small $x$. In the target rest frame, the electron-nucleus scattering at HERA allows a new regime to 
be probed experimentally for the first time. This is a new regime in 
which the virtual photon interacts coherently with all the nucleons at
a given impact parameter. This can 
be visualized in terms of the propagation of   a small $q\overline{q}$ pair
  in high density gluon fields through much larger distances than it is 
  possible with free nucleons.
Few years ago, a perturbative approach has been developed to calculate the gluon distribution in a nucleus \cite{ayala1} using perturbative QCD at small $x$.
This approach, known as  Glauber-Mueller (GM) approach is formulated in the target rest frame, takes into account the fluctuations of the hard probe.
It  includes the shadowing corrections (SC) due to parton rescatterings  inside the nucleus,
and provides the SC to the nuclear gluon distribution using the solution of the DGLAP evolution equations \cite{dglap} to the nucleon case.
As a result the behavior of related observables ($F_2^A, dF_2^A/dlogQ^2, F_L^A$, ...) at high energies can be calculated.

The GM approach was extended for the nucleon case in \cite{ayala2} and a comprehensive phenomenological analysis of the behavior of distinct observables ($F_2, F_L, F_2^c$) was made for the $ep$ HERA kinematical region using this approach \cite{ayala3}.
Our main conclusion was that the unitarity corrections are large in the HERA kinematical region, but only new data, with better statistics, will allow to  discriminate these corrections from the DGLAP predictions.
 The recent ZEUS data  \cite{df2zeus} for the slope of the proton structure function  presents a 'turn over' which cannot be reproduced by the DGLAP evolution equations with the GRV95 parameterization \cite{grv95}. Initially, this behavior was interpreted as the first evidence of the shadowing corrections in the  kinematic region of the $ep$ HERA collider \cite{glmn,vic2}. However, the MRST \cite{mrst} and GRV \cite{grv98} groups have produced a new set of parameterizations of the parton distributions which also reproduced the data.  Therefore the current situation  of the $ep$ HERA data still cannot demonstrate clearly the presence of the shadowing corrections. This conclusion motivates an analysis of these corrections in other processes. In this Letter we analyze  the $A$ dependence of the slope of the nuclear structure function which should be measured in the future in the $eA$ HERA collider.

Lets us start from the space-time picture of the $eA$ processes \cite{gribov}.
The deep inelastic scattering $eA \rightarrow e + X$ is characterized by a large electron energy loss $\nu$ (in the target rest frame) and an invariant momentum
transfer  $q^2 \equiv - Q^2$ between the incoming and outgoing electron 
such that $x = Q^2/2m_N \nu$ is fixed.  The general features of the time 
development  can be established using only
Lorentz invariance and the uncertainty principle. The incoming physical
electron state can, at a given instant of time, be expanded in terms of its
(bare) Fock states
\begin{eqnarray}
|e>_{phys} = \psi_e |e> + \psi_{e\gamma} |e \gamma> + ... \,\,.
\end{eqnarray}
The amplitudes $\psi_i$ depend on the kinematic variables describing the 
states $|i>$, and have the time dependence $exp(-iE_it)$, where $E_i = \sum_i 
\sqrt{m_i^2 + \vec{p}_i^{\,2}}$ is the energy of the state. The 'lifetime'
$\tau_i \approx 1/(E_i - E_e)$ of a Fock state $|i>$ is given by the time
interval after which the relative phase $exp[-i(E_i - E_e)]$ is significantly
different from unity. If $\tau_i > R_A$ the Fock state forms long before the 
electron arrives at the nucleus, and it lives long after its passage. 
New Fock states are not formed inside the nucleus. Therefore,
the scattering inside the nucleus is diagonal in the Fock basis. If the 
state $|i>$ contains particles with mass $m_j$, energy fraction $x_j$ and transverse
momentum $p_{t\,j}$, we have that the transverse velocities $v_{t\,j} =
p_{t\,j}/x_j E_e$ are small at large $E_e$. Hence the impact parameters (transverse
coordinates) of all particles are preserved. 

In terms of Fock states we then view the $eA$ scattering as follows: the 
electron emits a photon ($|e> \rightarrow |e\gamma>$) with $E_{\gamma} = \nu$
and $p_{t \, \gamma}^2 \approx Q^2$, after  the photon splits  into a $q\overline{q}$
($|e\gamma> \rightarrow |e q\overline{q}>$) and typically travels a  
distance $l_c \approx 1/m_N x$, referred as the 'coherence length', 
before interacting in the nucleus. For small $x$, the photon converts to a quark pair at a large distance before it interacts to the target; for example, at the $ep$ HERA collider, where one can study structure functions at $x \approx 10^{-5}$, the coherence length is as large as $10^4 \,fm$, much larger than the nuclear radii.
Consequently, the space-time picture of the DIS in the target
rest frame can be viewed as the decay of the virtual photon at high energy
(small $x$) into a quark-antiquark pair long before the 
interaction with the target. The $q\overline{q}$ pair subsequently interacts 
with the target.  In the small $x$ region, where 
$x \ll \frac{1}{2mR}$, the $q\overline{q}$  pair 
crosses the target with fixed
transverse distance $r_t$ between the quarks. It allows to factorize the total 
cross section between the wave function of the photon and the interaction 
cross section of the quark-antiquark pair with the target. The photon wave function 
is calculable and the interaction cross section is modelled. Therefore, the 
nuclear structure function is given by \cite{nik}
\begin{eqnarray}
F_2^A(x,Q^2) = \frac{Q^2}{4 \pi \alpha_{em}} \int dz \int \frac{d^2r_t}{\pi} |\Psi(z,r_t)|^2 \, \sigma^{q\overline{q} + A}(z,r_t)\,\,,
\label{f2target}
\end{eqnarray}
where 
\begin{eqnarray}
|\Psi(z,r_t)|^2 = \frac{6 \alpha_{em}}{(2 \pi)^2} \sum^{n_f}_i e_f^2 \{[z^2 
+ (1-z)^2] \epsilon^2\, K_1(\epsilon r_t)^2 + m_f^2\, K_0^2(\epsilon r_t)^2\}\,\,,
\label{wave}
\end{eqnarray}
$\alpha_{em}$ is the electromagnetic coupling constant,
$\epsilon = z(1-z)Q^2 + m_f^2$, $m_f$ is the quark mass, $n_f$ is the number 
of active flavors, $e_f^2$ is the square of the  parton charge (in units of $e$), $K_{0,1}$ 
are the modified Bessel functions and $z$ is the fraction of the photon's light-cone 
momentum carried by one of the quarks of the pair.  In the 
leading log$(1/x)$ approximation we can neglect the change of $z$ during the 
interaction and describe the cross section $\sigma^{q\overline{q}+A}(z,r_t^2)$ as 
a function of the variable $x$.

We estimated the unitarity corrections considering the  Glauber multiple scattering theory \cite{chou}, 
which was probed in QCD \cite{muegla}. The nuclear
 collision is analysed as a
sucession of collisions of the probe with individual nucleons within the nucleus, and summarizing we obtain that   
the $F_2$ structure function can be written  as \cite{ayala1,f2a}
\begin{eqnarray}
F_2^A(x,Q^2) =  \frac{R_A^2}{2\pi^2} \sum_1^{n_f} \epsilon_i^2 \int_{\frac{1}{Q^2}}^{\frac{1}{Q_0^2}} \frac{d^2r_t}{\pi r_t^4} \{C + ln(\kappa_q(x, r_t^2)) + E_1(\kappa_q(x, r_t^2))\}\,\,,
\label{diseik}
\end{eqnarray}
where $A$ is the number of nucleons, $R_A^2$ is the mean nuclear radius and $\kappa_q =  (2 \alpha_s\,A/3R_A^2)\,\pi\,r_t^2\,
 xG_N(x,\frac{1}{r_t^2})$ (see \cite{f2a} for details).

The slope of the nuclear structure function can be obtained directly from the expression (\ref{diseik}). We obtain that 
\begin{eqnarray}
\frac{d F_2^A(x,Q^2)}{dlog Q^2} =  \frac{R_A^2 Q^2}{2\pi^2} \sum_1^{n_f} \epsilon_i^2  \{C + ln(\kappa_q(x, Q^2)) + E_1(\kappa_q(x,Q^2))\}\,\,,
\label{diseik2}
\end{eqnarray}
which predicts the $x, \,Q^2$ and $A$ dependence of the shadowing corrections  for the $F_2^A$ slope. 

We see that the behavior of the  $F_2^A$ and its slope are strongly dependent on the nucleon gluon distribution. This is a common characteristic of the observables in the small $x$ region, where the gluon distribution dominates. 
Therefore, before we make predictions for these observables an analysis of the behavior of the gluon distribution should be made.
In the nucleon case,  a strong  growth of the nucleon structure function is observed in the $ep$ HERA data, which violates the unitarity boundary \cite{plb}. Consequently, we expect that unitarity corrections will be  present in the $ep$ HERA kinematical region.  
In \cite{vic2} we have shown that the ZEUS data for the $F_2^p$ slope \cite{df2zeus} which presents a turn over    can only  be successfully described considering the expression (\ref{diseik2}) for the nucleon case and a shadowed gluon distribution (quark + gluon sectors). In this Letter we consider that the nucleon gluon distribution is calculated using the GM approach, {\it i. e.} we consider that the behavior of the gluon distribution was modified by the unitarity corrections (see \cite{vic2} for details). 

The behavior of the nuclear structure function was analysed in \cite{f2a} using the Glauber-Mueller approach. We have shown that the ratio $R_1 = F_2^A/(A F_2^p)$ is strongly modified by the shadowing corrections and that it saturates in the perturbative regime ($Q^2 \ge 1\,GeV^2$) when both the quark and gluon sectors are considered. 
Here we estimate the shadowing corrections for the $F_2^A$ slope in the HERA kinematic region, where   $s = 9 \times 10^4 \,GeV^2$. Following \cite{df2zeus}, where the data points correspond to different $x$ and $Q^2$,  we consider that the variables $x$ and $Q^2$ are related by the expression $x = Q^2/(s y)$ and that  the inelasticity variable $y$ is given by $y = 0.25$. This is a typical value in the measurements at HERA \cite{h1y}. 

In figure (\ref{fig1}) we present our predictions  (solid curve) for the behavior of the $F_2^A$ slope using the expression (\ref{diseik2}).  We compare our results with the predictions of the DGLAP evolution equations using the GRV parameterization  (dashed curve) without any nuclear effect. We see that a 'turn over' is present in the DGLAP (GRV) and GM predictions.  However, the remarkable property of the result is that  the shadowing corrections modify the maximum for each of the slopes and that  it is $A$ dependent. This is expected from the formalism based on GM approach as extensively explained in \cite{ayala1}. In table (\ref{turn}) we present explicitly the $A$ dependence of the 'turn over' in the $F_2^A$ slope. 
In the Calcium case ($A = 40$) we predict that the 'turn over' occurs at $Q^2 = 5 \,GeV^2$, differently from the DGLAP (GRV) case for which is at $Q^2 = 1.7 \, GeV^2$  independently of $A$. We believe that this behavior cannot be mimicked by  modifications in the parton parameterizations, which makes $dF_2^A/dlogQ^2$ a sensitive probe of the shadowing corrections.

The behavior of the $F_2^A$ slope can be understand intuitively. The $A$ dependence of the 'turn over' is associated with the regime in which the partons in the nucleus form a dense system with mutual interactions and recombinations. The recombinations, {\it i.e} the shadowing corrections,  occur predominantly at large   density. As  the partonic density growth at larger values of the number of nucleons $A$ and smaller values of $x$, the same density at  $A = 1, 40, 197$ is obtained  at larger values of $x$ and in extension $Q^2$. This behavior of the recombinations is  verified in the $F_2^A$ slope.

Our main conclusion is that the analysis of the slope of the nuclear structure functions at $eA$ HERA energies will allow to discriminate the presence of the shadowing corrections from the DGLAP predictions. Our result  has important implications in the nucleon case and in  QCD at high densities. In the nucleon case, the evidence of an $A$ dependence of the 'turn over' will demonstrate that the correct way to estimate the observables in the $ep$ HERA collider is considering the shadowing corrections, without modifing the parton distributions. On the other hand,
in the near future, the collider facilities such as  
BNL Relativistic Heavy Ion Collider (RHIC), and CERN Large Hadron Collider 
(LHC) ($p\overline{p}$, $AA$) will be able to probe new regimes of dense 
quark matter at very small Bjorken $x$ or/and at large $A$, with rather 
different dynamical properties. The description of these processes  is 
directly associated with a correct description of the dynamics of minijet production, which will be strongly modified by shadowing corrections. 
We expect that this result contributes to motivate the running of nucleus at HERA in the future.

\section*{Acknowledgments}

This work was partially financed by CNPq and by Programa de Apoio a N\'ucleos de Excel\^encia (PRONEX), BRAZIL.

\newpage 
\section*{Tables}

\vspace{2.0cm}

\begin{table}[h]
\begin{center}
\begin{tabular} {||l|l|l|l|l||}
\hline
\hline
  & GRV & & GM &  \\
 & $x$ & $Q^2 \, (GeV^2)$ & $x$ & $Q^2 \, (GeV^2)$ \\
\hline
$A = 1$ & $0.75 \times 10^{-4}$ & $1.7$ & $0.48 \times 10^{-4}$  & 1.1 \\
\hline
$A = 40$ & $0.75 \times 10^{-4}$ & 1.7 & $0.22 \times 10^{-3}$ & 5.0 \\
\hline
$A = 197$ & $0.75 \times 10^{-4}$ & 1.7 & $0.28 \times 10^{-3}$ & 6.5 \\
\hline
\hline
\end{tabular}
\end{center}
\caption{The $A$ dependence of the 'turn over' predicted in the $F_2^A$ slope.}
\label{turn}
\end{table}

\newpage
\section*{Figure Captions}

\begin{figure}
\centerline{\psfig{file=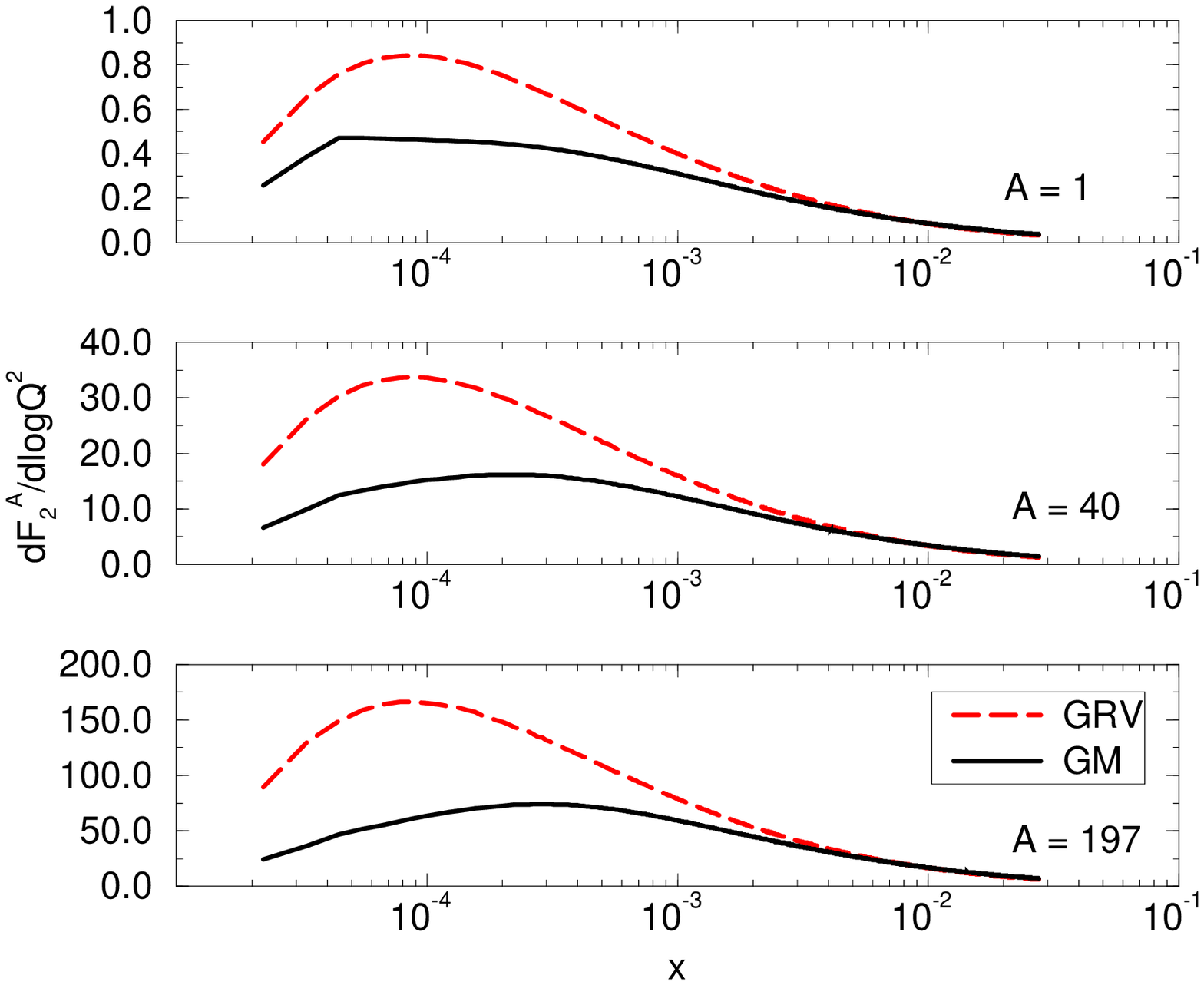,width=150mm}} 
\caption{ }
\label{fig1}
\end{figure}

\vspace{1.0cm}
Fig. \ref{fig1}: The $F_2^A$ slope as function of the variable $x$ at different values of $A$.  Each value of $x$ is related with the virtuality $Q^2$ by the expression $x = Q^2/ (s y)$, where we have assumed $s = 9 \times 10^4 \, GeV^2$ and $y = 0.25$. See text.

\newpage

\begin{thebibliography}{99}



\bibitem{arneodo}
New Muon Collab., M. Arneodo {\sl et al.} .{\sl Nucl. Phys.} {\bf B483} (1997) 3; { Nucl. Phys.} {\bf B441} (1995) 12; E665 Collab.,
M. R. Adams {\sl et al.} .  { \sl Z. Phys.} {\bf C67} (1995) 403.


\bibitem{hera96} 
M. Arneodo {\sl et al.}.  {\sl Future Physics at HERA.} Proceedings of the Workshop 1995/1996. Edited by G. Ingelman {\sl et al.}.


\bibitem{qiu}
L. V. Gribov, E. M. Levin, M. G. Ryskin. { \sl Phys. Rep.} {\bf 100} (1983) 1; A. H. Mueller, J. Qiu.  { \sl Nucl. Phys.} {\bf B268} (1986) 427.

\bibitem{ayala1}
A. L. Ayala, M. B. Gay Ducati and E. M. Levin.  {\sl Nucl. Phys.} {\bf B493} (1997) 305.

\bibitem{dglap}

 Yu. L. Dokshitzer. { Sov. Phys. JETP} {\bf 46} (1977) 641;
 G. Altarelli and G. Parisi. { Nucl. Phys.} {\bf B126} (1977)  298;
 V. N. Gribov and L.N. Lipatov. { Sov. J. Nucl. Phys} {\bf 15} (1972) 822.

\bibitem{ayala2}
A. L. Ayala, M. B. Gay Ducati and E. M. Levin.  
 {\sl Nucl. Phys.} {\bf B511} (1998) 355.
 
\bibitem{ayala3}
A. L. Ayala, M. B. Gay Ducati and E. M. Levin. {\sl Eur. Phys. J.} {\bf C8} (1999) 115; 
A.L. Ayala, M. B. Gay Ducati, V. P. Gon\c{c}alves.  {\sl Phys. Rev.} {\bf D59} (1999) 054010.


\bibitem{df2zeus}
Zeus Collab., M. Derrick {\it et al.} . {\sl Eur. Phys. J.}  {\bf C7} (1999) 609.



\bibitem{grv95} 
M. Gluck, E. Reya and A. Vogt.  { \sl Z. Phys.} {\bf C67} (1995) 433.

\bibitem{glmn}
E. Gotsman {\it et al.} . {\sl  Nucl. Phys.} {\bf B539} (1999) 535.







\bibitem{vic2}
M. B. Gay Ducati and V. P. Gon\c{c}alves. SLAC-PUB-7968 (October 1998), hep-ph/9812459.




\bibitem{mrst}
A. D. Martin {\it et al.}. {\sl Eur. Phys. J.}  {\bf C4} (1998) 463.


\bibitem{grv98} 
M. Gluck, E. Reya and A. Vogt.  {\sl Eur. Phys. J.}  {\bf C5} (1998) 461.


\bibitem{gribov}
V. N. Gribov.  {\sl Sov. Phys. JETP} {\bf 29} (1969) 483;
V. Del Duca, S. J. Brodsky and P. Hoyer. {\sl Phys. Rev.} {\bf D46} (1992) 931.


\bibitem{nik} 
N. Nikolaev and B. G. Zakharov,  {\sl Z. Phys.} {\bf C49}  (1990) 607.


\bibitem{chou}
R. J. Glauber, {\sl  Phys. Rev.} {\bf 100} (1995) 242; R. C. Arnold, {\sl  Phys. Rev.} {\bf 153} (1967) 1523;
T. T. Chou and C. N. Yang, {\sl  Phys. Rev.} {\bf 170} (1968) 1591.


\bibitem{muegla}
A. H. Mueller.  { Nucl. Phys.} {\bf B335} (1990) 335.



\bibitem{f2a}
M. B. Gay Ducati and V. P. Gon\c{c}alves.  hep-ph/9902236.


\bibitem{plb}
A. L. Ayala, M. B. Gay Ducati and E. M. Levin. {\sl Phys. Lett.} {\bf B388} (1996) 188.


\bibitem{h1y}
H1 Collab., S. Aid {\it et al.} . { Nucl. Phys.} {\bf B370} (1996) 3.








 



\end{thebibliography}
\end{document}